\documentclass{svjour3}                     
\smartqed  
\usepackage{graphicx}
\begin{document}

\title{Coupled channel approach to (S,I)=(-2,0) baryon-baryon interactions from lattice QCD
\thanks{Presented at the 21st European Conference on Few-Body Problems in Physics, Salamanca, Spain, 30 August - 3 September 2010}
}


\author{K. Sasaki   \hspace*{.5em} 
        for HAL QCD collaboration. 
        \scalebox{.5}{\includegraphics{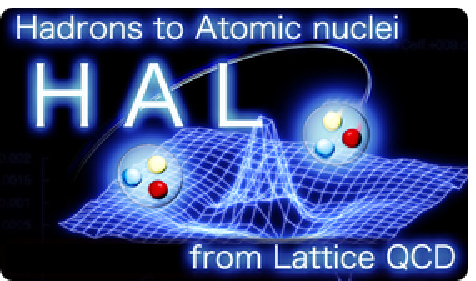}}
}


\institute{K. Sasaki \at
              Graduate School of Pure and Applied Sciences, University of Tsukuba, Japan \\
              \email{kenjis@het.ph.tsukuba.ac.jp}           
}

\date{Received: date / Accepted: date}

\maketitle

\begin{abstract}
We investigate baryon-baryon interactions with strangeness $S=-2$ and isospin $I=0$ system from Lattice QCD.
In order to solve this system, we prepare three types of baryon-baryon operators ($\Lambda\Lambda$, $N\Xi$ and $\Sigma\Sigma$) for the sink and construct three source operators diagonalizing the $3\times3$ correlation matrix.
Combining of the prepared sink operators with the diagonalized source operators, we obtain nine effective Nambu-Bethe-Salpeter (NBS) wave functions.
The $3\times3$ potential matrix is calculated by solving the coupled-channel Schr\"odinger equation.
The flavor $SU(3)$ breaking effects of the potential matrix are also discussed by comparing with the results of the $SU(3)$ limit calculation.
Our numerical results are obtained from three sets of $2+1$ flavor QCD gauge configurations provided by the CP-PACS/JLQCD Collaborations.
\end{abstract}

\section{Introduction}
Completion of the knowledge of the generalized nuclear force, which includes not only the nucleon-nucleon (NN) interaction but also hyperon-nucleon (YN) and hyperon-hyperon (YY) interactions, brought the deeper understanding of atomic nuclei, structure of neutron stars and supernova explosions.
However it is hard to know the properties of the YN and YY interactions because their scattering data in free-space are scarce.

Recently a method to extract the $NN$ potential through the NBS wave function from lattice QCD simulations has been proposed in~\cite{Ishii:2006ec}.
The obtained potential is found to have desirable features, such as attractive well at long and medium distances, and the central repulsive core at short distance~\cite{Ishii:2006ec,Aoki:2009ji}. 
Further applications have been done in Refs.~\cite{Nemura:2008sp,Nemura:2009kc,Murano:2010tc,Ikeda:2010sg,Inoue:2010hs}.

In this work, we focus on the $S=-2$, $I=0$ B-B system to seek the $\Lambda \Lambda$ interaction and to see the SU(3)$_f$ breaking effects of B-B interaction from lattice QCD simulation.

\section{Formulation}
The $(S,I)=(-2,0)$ baryon-baryon state consists of the $\Lambda \Lambda$, $N \Xi$ and $\Sigma \Sigma$ components in terms of low-lying baryons.
Mass differences of these components are quite small, and it causes the contamination of NBS wave function from excited states.
In sucn situation the source operator should be optimized to extract the energy eigen states through the variational method~\cite{Michael:1985ne,Luscher:1990ck}.

The equal-time NBS wave function $\psi^{B_1B_2}(\vec{r},E)$ for an energy eigen state with $E$ is extracted from the four point function,
\begin{equation}
 W^{B_1B_2}(t-t_{0},\vec r)
= \sum_{\vec x} \langle 0 \mid B_1(t,\vec x+\vec r)\,B_2(t,\vec x)\,
                       {\bar {\cal{I}}_E}(t_0) \mid 0 \rangle ~,
\end{equation}
where ${\cal{I}}_E$ is diagonalized wall-source operator.
The transition potential matrix of 3-states coupled channel equation can be acquired in a particle basis or a SU(3) irreducible representation (IR) basis.
They are connected by unitary trandformation (see in Appendix B in Ref.~\cite{Inoue:2010hs}).
The non-diagonal part of potential matrix in IR basis is a good measure of the SU(3) breaking effect.

\section{Numerical setup}
\begin{table}
\begin{center}
\caption{Hadron masses in unit of [MeV] are listed.}
 \label{TAB:Gconf}
  \begin{tabular}{cccccccc}
  \hline \hline
   & $\kappa_{ud}$ & $m_\pi$ & $m_K$ & $m_N$ & $m_\Lambda$ & $m_\Sigma$ & $m_\Xi$ \\
  \hline
  Set 1 & $0.13760$ & $875(1)$ & $916(1)$ & $1806(3)$ & $1835(3)$ & $1841(3)$ & $1867(2)$ \\
  Set 2 & $0.13800$ & $749(1)$ & $828(1)$ & $1616(3)$ & $1671(2)$ & $1685(2)$ & $1734(2)$ \\
  Set 3 & $0.13825$ & $661(1)$ & $768(1)$ & $1482(3)$ & $1557(3)$ & $1576(3)$ & $1640(3)$ \\
  \hline \hline
  \end{tabular}
 \end{center}
\end{table}
In this calculation we employ the 2+1-flavor full QCD gauge configurations of Japan Lattice Data Grid(JLDG)/International Lattice Data Grid(ILDG).
They are generated by the CP-PACS and JLQCD Collaborations with a renormalization-group improved gauge action and a non-perturbatively $O(a)$ improved clover quark action at $\beta = 6/g^2=1.83$, corresponding to lattice spacings of $a = 0.1209 {\rm{fm}}$~\cite{Ishikawa:2007nn}.
We choose three ensembles of the $L^3 \times T = 16^3 \times 32$ lattice which means the spatial volume of about $(2.0 {\rm{fm}})^3$. 
Quark propagators are calculated from the spatial wall source at $t_0$ with the Dirichlet boundary condition in temporal direction at $t - t_0 = 16$. 
The numerical computation is carried out at KEK supercomputer system, Blue Gene/L.
The hadron masses are shown in Table~\ref{TAB:Gconf}.

\section{Result}
\begin{figure}
 \begin{center}
  \rotatebox[origin=tl]{90}{\hspace*{3em}\tiny{V[MeV]}}
  \scalebox{0.38}[0.23]{\includegraphics{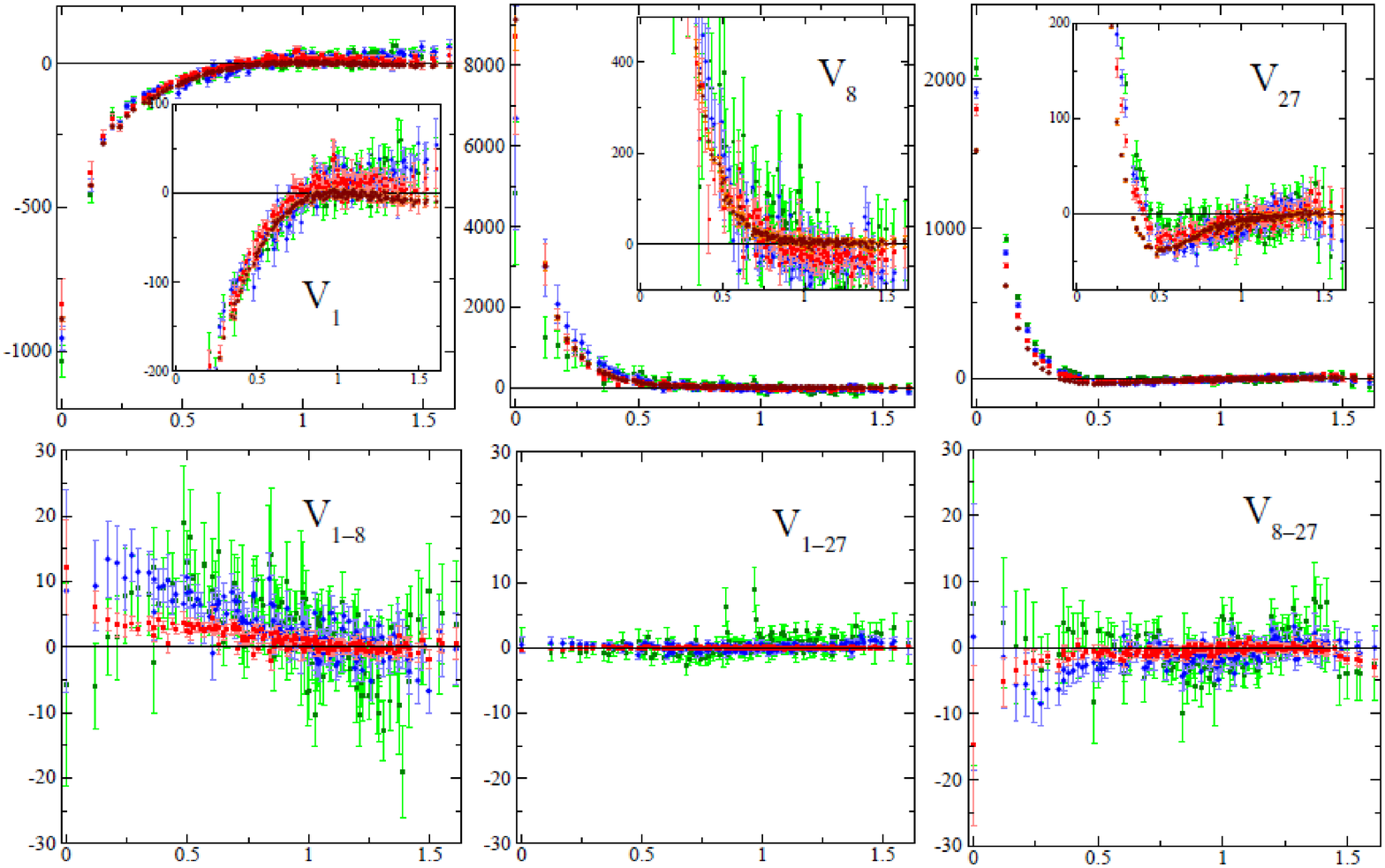}}\\[-0.2em]
  {\tiny{r[fm]}}
 \end{center}
\caption{Transition potentials in the SU(3) IR basis. Red, blue and green symbol respectively correspond to the result of Set1, Set2 and Set3. The result of the SU(3) symmetric configuration is also plotted with brown symbol.}
\label{FIG:potall}
\end{figure}

In Figure~\ref{FIG:potall} we compare the results of potential matrix in the IR basis calculated in different configuration sets.
We found the growth of repulstive core in the $V^{27}$ potential with decreasing the light quark mass.
The $V_{1-27}$ and $V_{8-27}$ transition potential are consistent with zero within error bar.
On the other hand, it is noteworthy that the $V_{1-8}$ transition potential which is not allowed in the SU(3) symmetric world is strengthen as the SU(3)$_f$ breaking gets larger.

\section{Conclusion}
We have investigated the $(S,I) = (-2,0)$ BB state, which is known as the $\Lambda \Lambda$, $N \Xi$ and $\Sigma \Sigma$ coupled state, from lattice QCD.
We have found a small transition potential between the singlet and octet state in terms of the SU(3) IR basis.
Such transition can not be allowed in the SU(3) symmetric world.
This method could greatly assist us to complete the knowledge of not only the generalized nuclear force but also the interaction of hadrons including mesons, baryons and quarks.

\vspace*{1em}
\noindent{\bf{Acknowledgements}}:
This work was supported by the Large Scale Simulation Program No.0923(FY2009) of High Energy Accelerator Research Organization (KEK), Grant-in-Aid of the Ministry of Education, Science and Technology, Sports and Culture (Nos. 20340047, 22540268, 19540261) and the Grant-in-Aid for Scientific Research on Innovative Areas (No. 2004:20105001, 20105003).

\end{document}